\documentclass[traditabstract]{aa}
\usepackage{graphicx}
\usepackage{txfonts}
\usepackage{natbib}
\usepackage{ulem}
\usepackage{color}

\begin{document}
\bibliographystyle{aa}

   \title{The impact of X-rays on molecular cloud fragmentation and the IMF}
   \subtitle{}

   \author{S. Hocuk 
          \and
          M. Spaans 
          }
   \institute{Kapteyn Astronomical Institute, University of Groningen,
              P.~O.~Box 800, 9700 AV Groningen \\
              \email{seyit@astro.rug.nl, spaans@astro.rug.nl}
             }

   \titlerunning{The impact of X-rays on molecular cloud fragmentation}
   \authorrunning{S.~Hocuk \& M.~Spaans}
   \date{Received \today}

\abstract
{Star formation is regulated through a variety of feedback processes. In this study, we treat feedback by X-rays and discuss its implications. Our aim is to investigate whether star formation is significantly affected when a star forming cloud resides in the vicinity of a strong X-ray source. We perform an Eulerian grid simulation with embedded Lagrangian sink particles of a collapsing molecular cloud near a massive, 10$^{7}$ M$_{\odot}$ black hole. The chemical and thermal changes caused by radiation are incorporated into the FLASH code. When there is strong X-ray feedback the star forming cloud fragments into larger clumps whereby fewer but more massive protostellar cores are formed. Competitive accretion has a strong impact on the mass function and a near-flat, non-Salpeter IMF results.}

\maketitle

\section{Introduction}
At a distance of 100 pc, a massive $\rm 10^7 ~ M_{\odot}$ black hole is able to emit X-rays with a flux of $\rm 100 ~erg ~s^{-1} ~cm^{-2}$ \citep{2007A&A...461..793M}. The presence of X-rays in star forming regions alters fundamental processes in star forming molecular clouds, that eventually determine stellar masses, like ionization and gas and dust heating, and do so up to column densities of $\rm 10^{24} ~cm^{-2}$ and distances of 300 pc \citep{2010A&A...513A...7S}.

In our local neighborhood, the initial mass function (IMF) is observed to be a power-law, nicely following a Salpeter slope. The IMF is assumed to have a universal shape and spatially well resolved studies \citep{2003PASP..115..763C, 2007AAS...210.1601S, 2008ApJ...681..365E} confirm this at low and high masses. When we turn to observations of more radical environments like the Galaxy center, hints for different IMFs appear \citep{2005Natur.434..192F, 2005ASSL..327...89F, 2006ApJ...643.1011P, 2009A&A...501..563E, 2009eimw.confE..14E, 2010ApJ...708..834B}. However, the evidence is also claimed to be controversial \citep{2010arXiv1001.2965B}. Nuclei of active galaxies, e.g., ULIRGs like Arp 220 and Markarian 231, see \cite{2010A&A...518L..42V}, enjoy conditions much more extreme than the Milky Way galactic center. Little is known on the IMF there, so theory and simulations are needed to guide understanding. The idea arose that massive stars form more easily in these extreme regions and that the IMF might be top-heavy \citep{2007MNRAS.377.1439B, 2007MNRAS.374L..29K, 2009MNRAS.394.1529D}. Other explanations are possible, still, it is worthwile to test this hypothesis numerically.

Until now, there have been many studies of feedback processes and external influences on star formation \citep{2008Sci...321.1060B, 2008ApJ...675..188W, 2009ApJ...702...63W, 2010ApJ...713.1120K, 2010MNRAS.404L..79B}. However, no numerical simulations have been performed yet of a collapsing molecular cloud under strong incident X-ray radiation. In this paper, we present the first 3D numerical simulations on the effects of X-rays on star forming molecular clouds that reside close to a massive black hole. In the next sections, we compare such a molecular cloud to an X-ray free environment and evaluate the star formation efficiency, stellar masses, and the resulting IMF.

\section{Numerical model}
\subsection{The FLASH code}
All simulations in this study have been performed with the Eulerian hydrodynamical and N-body code FLASH \citep{2000ApJS..131..273F, Dubey2009512}. We used FLASH with the improved gravity solver that was added in version 3 \citep{2008ApJS..176..293R}. FLASH is a modular based, strongly-scaled parallel code that is specialized in adaptively refined meshes in which one can use particles in conjunction with the grid.

The hydrodynamic equations are solved using the piecewise parabolic method \citep[PPM,][]{1984JCoPh..54..174C}, which is an improved version of Godunov's method \citep{Godunov}. PPM is particularly well suited for flows involving discontinuities, such as shocks, that are strongly present in this study. FLASH is provided with many and extensively tested modules that encompass a broad range of physics. Many of these modules, including hydrodynamics, thermodynamics, (self-)gravity, particles, turbulence, and shocks are standard ingredients for interstellar physics and star formation and are thus incorporated in this work. Several additions were made to the code to follow the (radiation)physics in detail, such as sink particles, radiative transfer, multi-scale turbulence, and refinement criteria based on Jeans length and sink particles. The main additions are described in the following subsections.

\subsection{Sink particles}
Sink particles are point particles that can grow in mass by accreting gas and merge with other particles, however, they cannot lose mass or fragment. These particles represent compact objects, in our case protostars. Sink particles are a necessary ingredient if one wants to follow a density evolution over a large dynamic range, like a collapsing cloud. We created a sink particle module that takes care of the particle creation, determines their masses, handles Bondi-Hoyle type accretion, and tracks the particles if they are eligible for merging.

A sink particle is created when it passes several criteria for indefinite collapse as given by \cite{2010ApJ...713..269F}. The mass it obtains at creation is determined by a density threshold that is set by the maximum resolution. The \cite{1997ApJ...489L.179T} criterion states that the Jeans length should not be resolved by less then 4 cells if one wants to avoid artificial fragmentation. This criterion can be rewritten in the form of a density limit. Any excess density is taken away from the grid and added to the mass of the particle. The sink particle algorithm created for this study follows the method created by \cite{2004ApJ...611..399K}. For any details of this method, we refer the reader to that paper.

\subsection{Radiative transfer}
In order to update the effects caused by X-rays on the temperature of the gas, we ported an X-ray dominated region chemical code (XDR code) created by \cite{2005A&A...436..397M} into FLASH. This code incorporates all of the heating (photo-ionization, yielding non-thermal electrons) and cooling processes from atomic (fine-structure, semi-forbidden) and molecular transitions (CO, H$_{2}$, H$_{2}$O). Effects from internal UV, cosmic rays, and dust-gas coupling are treated as well. Given an X-ray flux, gas density, and column density along the line of sight to the source, the XDR code calculates the temperature and the chemical abundances. This output is fed into the simulation at every iteration. Most of the computation is spent finding the column densities for every cell. A ray-tracing algorithm, specifically created for this purpose, searches the grid and sums up the column densities of each cell lying along the line of sight from the source, the accreting black hole, to the target cell. The X-ray flux is an E$^{-0.9}$ power law between 1 and 100 keV. X-ray scattering is not very important, but is nonetheless treated in the XDR-code. A uniform background of cosmic rays prevents the temperature from dropping below 10 K. For this, a cosmic ray ionization rate typical for the Milky Way $\rm \zeta=5\times10^{-17} ~s^{-1}$, is assumed \citep{2005ApJ...626..644S}

\begin{figure}[htb!]
\centering
\includegraphics[scale=0.33]{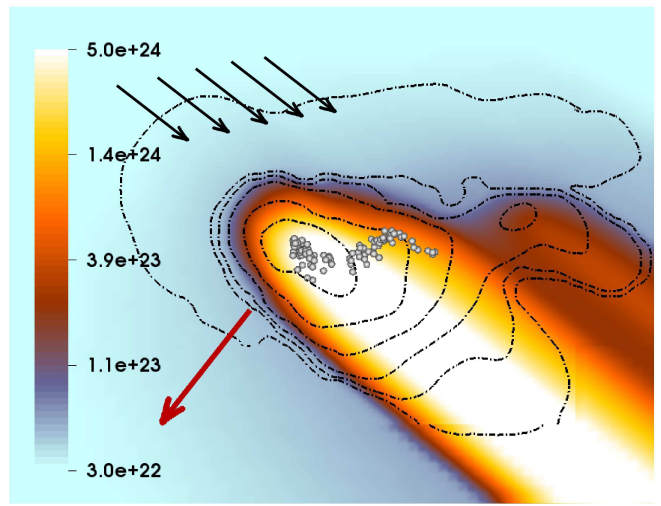}
\caption{Column density ($\rm cm^{-2}$) slice of simulation A at $\rm t=t_{ff}$. The colors represent the column density along the line of sight to the black hole, which is located at the upper left side (the black arrows indicate the direction of radiation). The density is shown in black contours. Contour levels are 1, 4, 16, 64, 256, 1024 $\rm \times 10^{4} ~cm^{-3}$. The white spheres display the location of the sink particles and the red arrow indicates the direction of the cloud's orbital motion.}
\label{fig:column}
\end{figure}

\subsection{Initial conditions}
We create a spherical gas cloud with solar abundances at a distance of 10 parsec from a 10$^7 \rm M_{\odot}$ black hole. We run two separate simulations and name them simulation A and B. Simulation A represents a molecular cloud near an active black hole under the impact of X-rays. Simulation B has a cloud near an inactive black hole and has isothermal conditions, with an equation of state of the form $\rm P \propto \rho$. The XDR code that updates the thermodynamics is only linked with simulation A where we do have X-rays. The 10$^7 \rm M_{\odot}$ black hole, at 10\% of Eddington, yields a flux of 160 $\rm erg ~s^{-1} ~cm^{-2}$ with some extinction \citep{2007A&A...461..793M}. \cite{2009ApJ...702...63W} show that column densities of $\rm 10^{22.5} cm^{-2}$ can typically exist in the central $\rm R \simeq 10 ~pc$ of an AGN, leading to a 1 keV optical depth of 2-3. Furthermore, \cite{2009ApJ...702...63W} show that column densities as large as 10$^{24} \rm ~cm^{-2}$ occur and persist in a statistical sense in the dynamically active inner 20 pc. A gaseous cloud can thus be shielded by this clumpy medium around the AGN and remain cold, while the temperature rises rapidly once the cloud is exposed to the radiation. Both our simulations start with the same initial conditions, but we expose simulation A to an X-ray source once the simulation starts. In this, heating is nearly immediate, since the timescale for heating is much shorter than the collapse time, $\rm t_{heat} \ll t_{ff}$, and of the order of 10$^{-1}$ years. All other conditions are the same for both simulations.

The simulations are set up with an initial random, divergence-free turbulent velocity field and a characteristic FWHM of 5 km/s that agrees well with molecular clouds observed in active regions \citep{2009A&A...503..459P}. These are supersonic flows with Mach numbers of up to 25, where the isothermal sound speed of the cloud is $\rm c_{s}=0.19$ km/s (for T=10K) and can go up to a maximum of 5 km/s when the cloud is heated by X-rays (T$\simeq$10$^{4}$ K). We do not drive the turbulence but follow its decay. The turbulence is applied over all scales with a power spectrum of $\rm P(k) \propto k^{-4}$, following the empirical laws for compressible fluids \citep{1981MNRAS.194..809L, 1999ApJ...522L.141M, 2004ApJ...615L..45H}. We start the simulations with a cloud that is in a stable Keplerian orbit around the black hole. Shear that is introduced by the black hole is taken into account. The maximum velocity difference imposed by the black hole, $\triangle \rm v_{shear}=$ 2.2 km/s across the cloud, is of the order of the applied initial turbulence. This process keeps the turbulence strong to large dynamical times, i.e., $\rm >1 ~t_{ff}$, with $\rm t_{ff} = \sqrt{ 3\pi / \rm 32G\rho}$ and 10$^{5}$ years\footnote{Given the initial condition of $\rho = 3.84 \cdot 10^{-19} \rm g/cm^{-3}$} throughout this work. The shearing time, $\rm t_{shear}=D_{cloud}/\triangle \rm v_{shear}$, is almost 3 times larger than the cloud free-fall time and gravitationally bound (roughly) spherical clouds are likely to exist at densities of $\sim$10$^{5} \rm ~cm^{-3}$.

\begin{figure*}[htb!]
\flushleft
\begin{tabular}{c c c}

\begin{minipage}{7.8cm}
\includegraphics[scale=0.48]{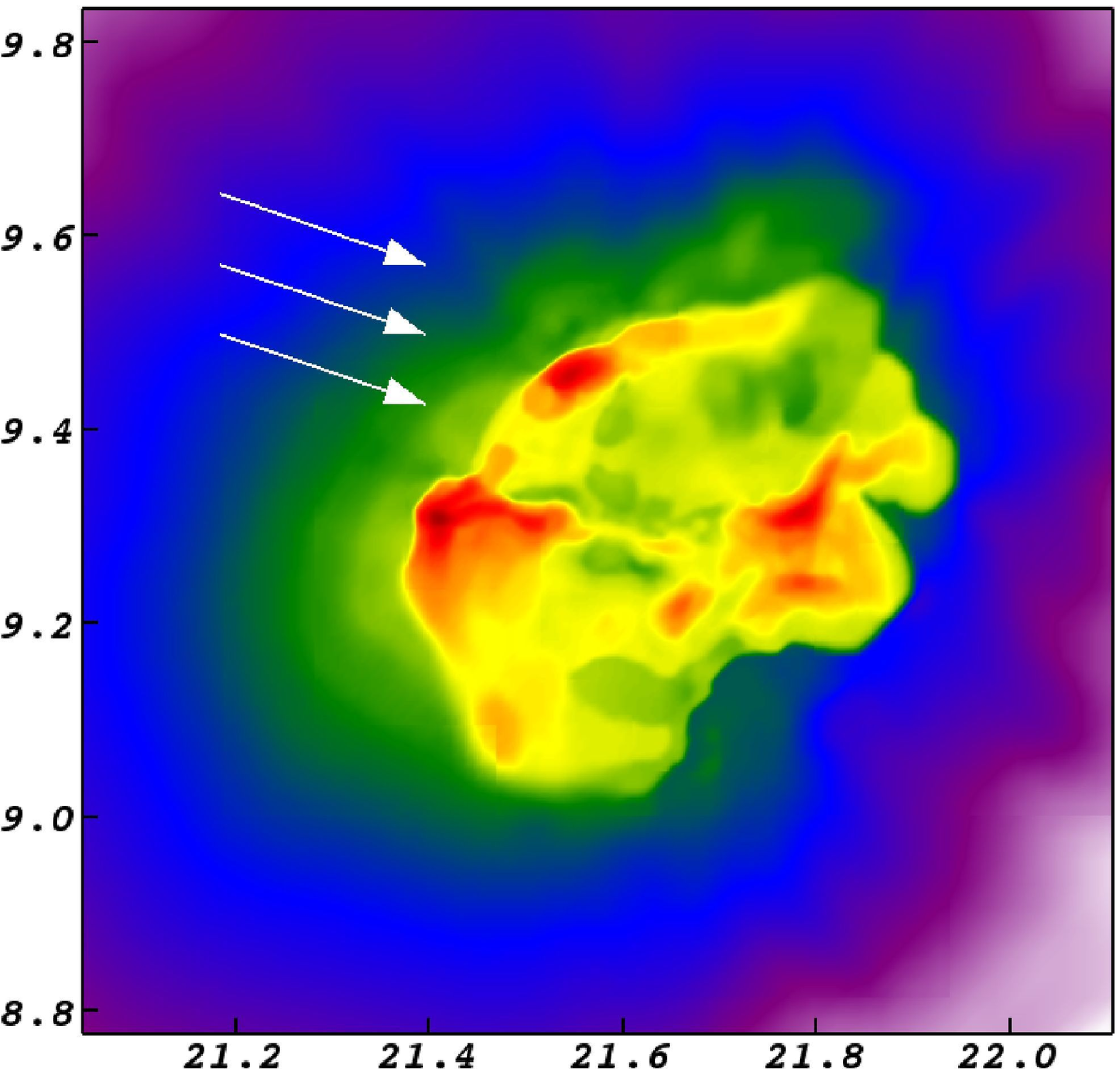}
\end{minipage} &

\begin{minipage}{8cm}
\includegraphics[scale=0.48]{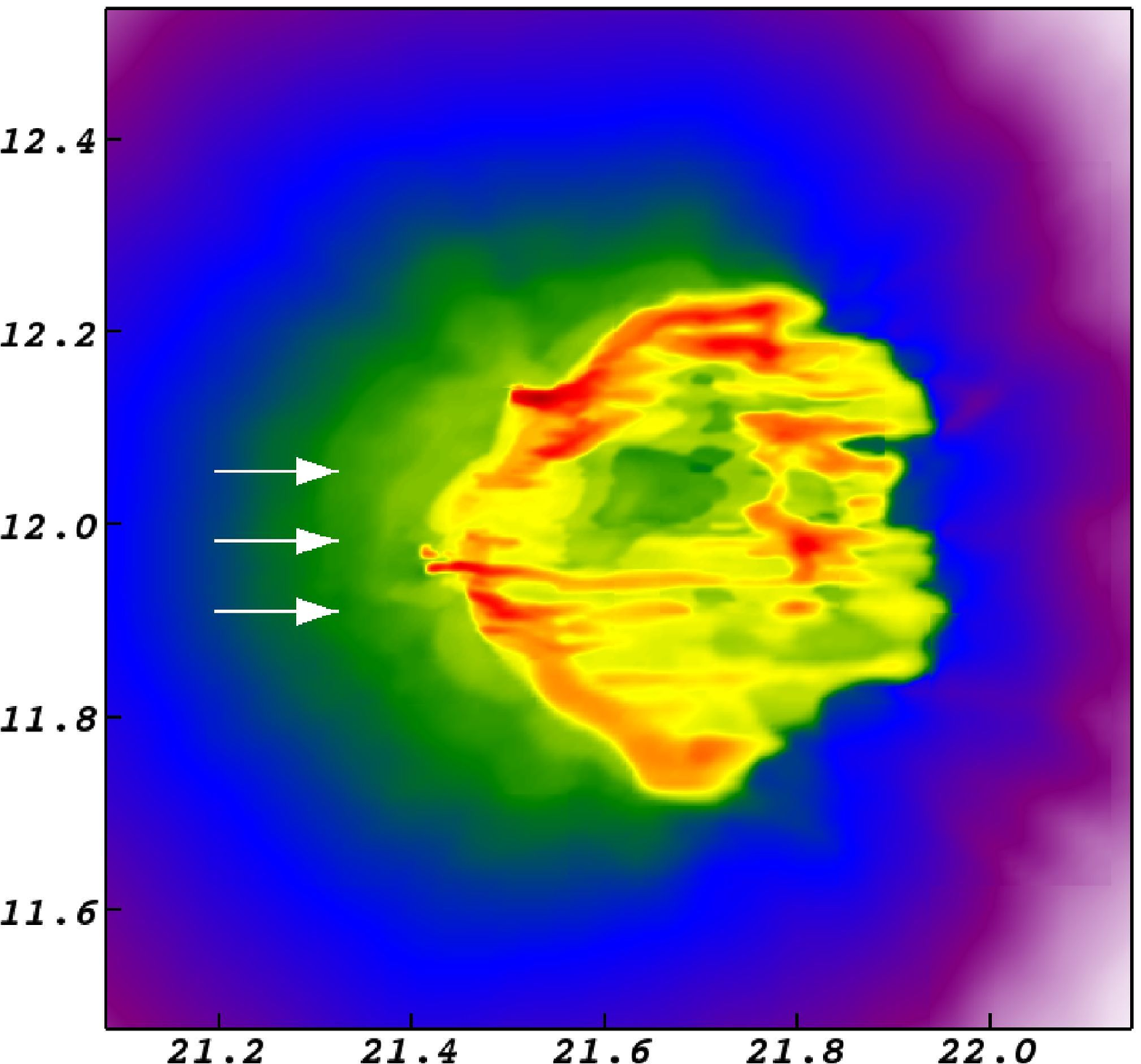}
\end{minipage} &

\begin{minipage}{0cm}
\includegraphics[scale=0.64]{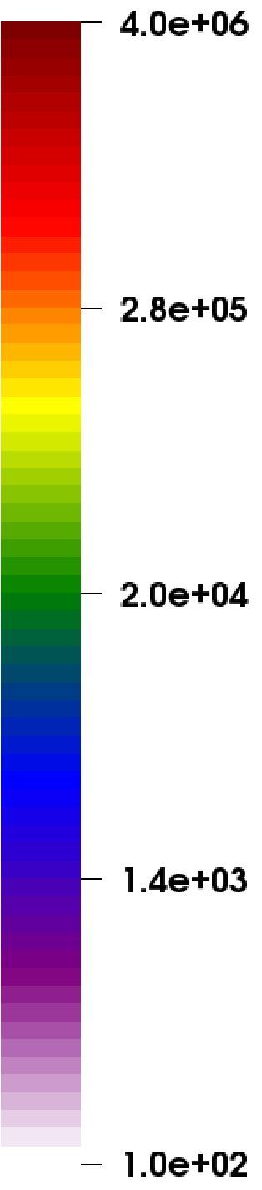}
\end{minipage} \\

\end{tabular}
\caption{Density morphology of simulation A at t=0.4t$_{\rm ff}$ in 2D slices through the center of the cloud with axes in parsec. The color represents the number density (cm$^{-3}$). On the left, an XY slice is shown with arrows representing the direction of radiation emanating from the black hole, which is located at the upper left side. On the right, an XZ slice is shown with radiation that is coming from the left and under an angle of 16 degrees with the slice.}
\label{fig:gaps}
\end{figure*}

The molecular cloud starts with a uniform number density of $10^{5} \rm ~cm^{-3}$ and has a size of 0.33 pc in radius. With a mean molecular weight of $\mu$ = 2.3, the total mass of the cloud amounts to 800 solar masses. The rest of the medium is filled with gas that has a uniform density of 100 $\rm cm^{-3}$. The simulation box, a cube of size 24 pc, has outflow boundaries and is isolated in terms of gravity. We increase the resolution where needed according to a self-developed Jeans criterion. The algorithm calculates the Jeans length at every grid cell, compares it against the Truelove criterion, and adds resolution when this is about to be violated. The maximum grid resolution that we allow for any simulation is 8192$^{3}$ cells. With the box size of 24 pc, the maximum spatial resolution becomes $8.8 \times 10^{15}$ cm. If gas continues to collapse and needs higher resolution beyond the maximum refinement level, sink particles are created thereby taking density away from the gas such that mass and momentum is conserved and the resolution criterion is not violated. We give sink particles an accretion radius of effectively 3 cells, that is, $2.6 \times 10^{16}$ cm $\simeq$ 1760 AU. The temperature in both simulations is initialized at 10 K, but for the simulation with X-rays, the temperature is updated by the XDR code, and changes quickly ($\lesssim$10$^{-4} \rm ~t_{ff}$ = 1 timestep) after initialization.

\section{Results}
\subsection{X-rays versus no X-rays: IMF and Jeans mass}
Both simulations, A and B, are followed for 2$\times$10$^{5}$ years. This is approximately two free-fall times for a $10^{5} \rm ~cm^{-3}$ cloud. X-rays can heat the cloud to as high as 6000 K at low column densities ($< 10^{21} \rm ~cm^{-2}$), this is the case for the side that is directly exposed to the X-ray source, but also to as low as 10 K at high column densities ($> 10^{24} \rm ~cm^{-2}$). Fig. \ref{fig:column} shows a column density plot to this effect.

\begin{figure}[htb!]
\centering
\includegraphics[scale=0.51]{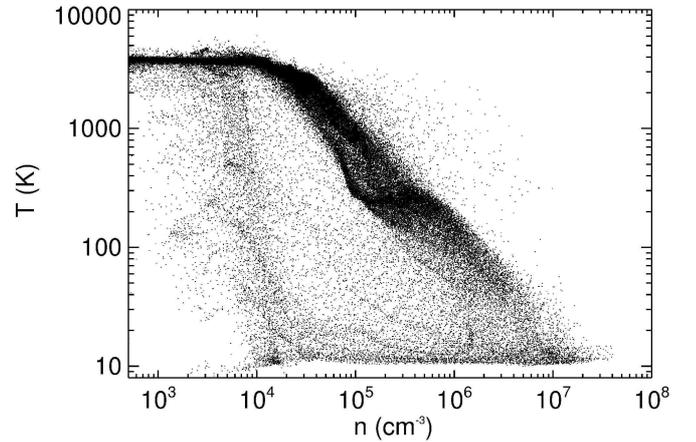}
\caption{Temperature-density phase diagram of the X-ray irradiated simulation (A) after 1 dynamical evolution, $\rm t=t_{ff}$.}
\label{fig:phase}
\end{figure}

The directional heating increases the pressure and causes the gas to expand and evaporate on the irradiated face of the cloud. The molecular cloud is compressed, loses mass, and an ionizing pressure flow travels inward. We see that this compression creates a density increase of about half an order of magnitude within a free-fall time as compared to simulation B. The conical compression front is disrupted where the turbulence creates sub-pc scale gaps (0.01-0.05 pc) and radiation is able to penetrate, as is illustrated in Fig. \ref{fig:gaps}. Those regions are also heated up and pressurized. This causes the pressure front of the irradiated side of the cloud to break up. We see finger-like shapes forming, with a high density head, and the gas that is lying in its shadow is well shielded and very cold (about 10 K). The increased density induces star formation. We find that sink particles are created in the compressed cloud edge much earlier than in the shielded and colder parts. A phase diagram is plotted in Fig. \ref{fig:phase} that shows the decline of temperature as the density increases. The secondary band with a steep decrease in temperature at low densities is the direct result of shielding. We note that the reason that fewer points appear in the plot in the shielded regions is merely due to the difference in resolution determined by the refinement criteria.

\begin{figure*}[htb!]
\flushleft
\begin{tabular}{l l}
\begin{minipage}{8cm}
\hspace{-1cm}
\includegraphics[scale=0.55]{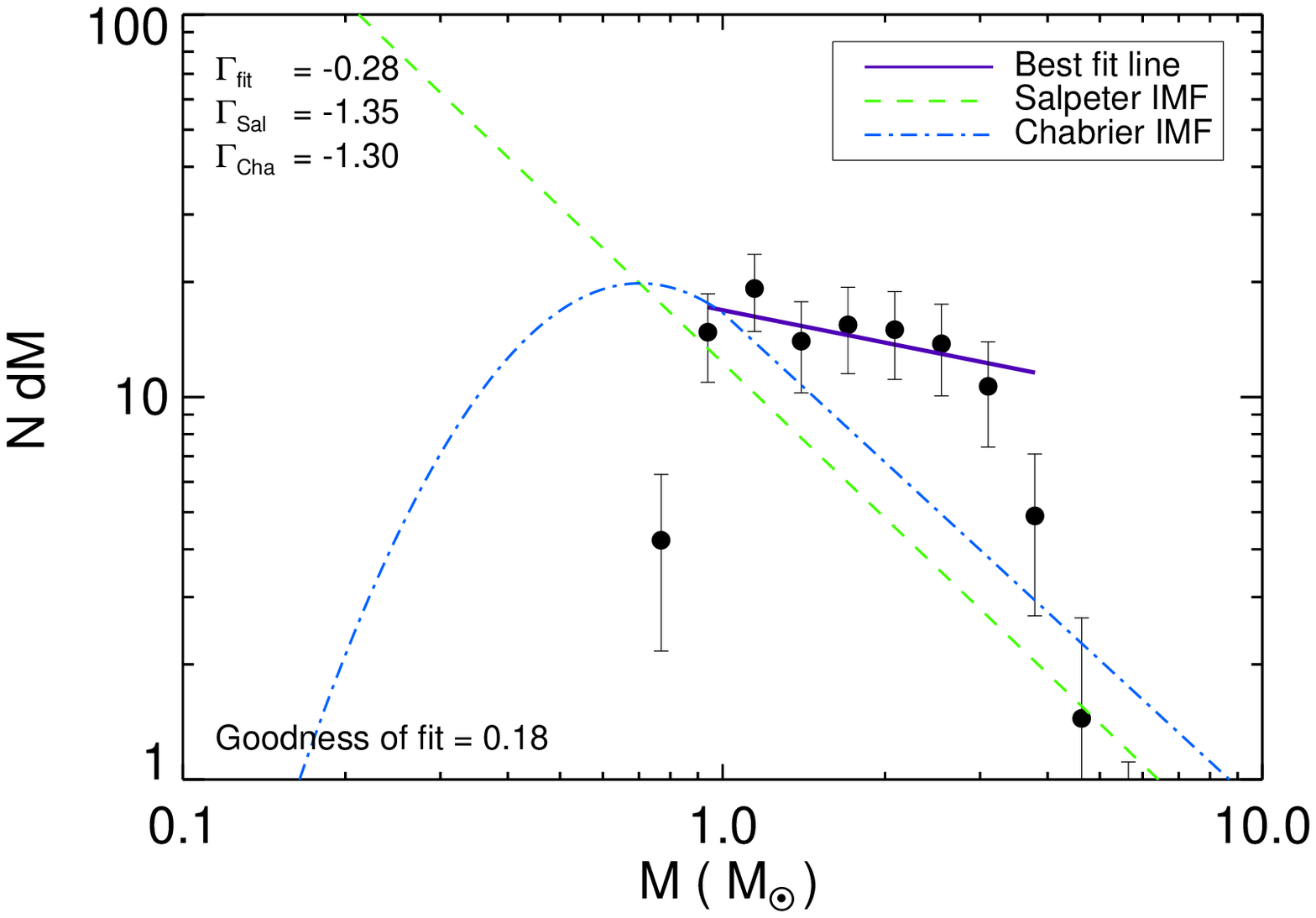}
\end{minipage} &

\begin{minipage}{0cm}
\includegraphics[scale=0.55]{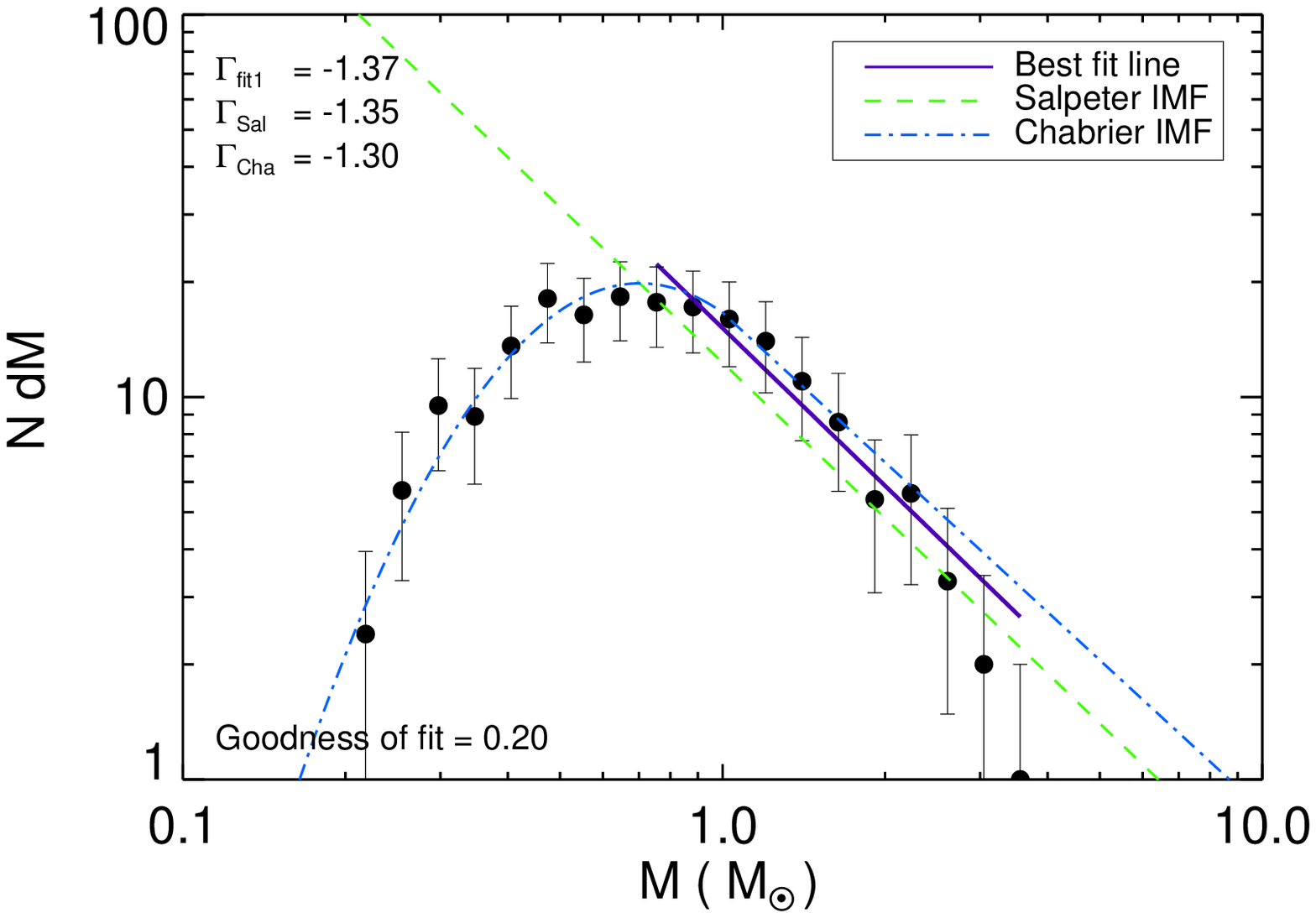}
\end{minipage} \\

\end{tabular}
\caption{The initial mass function of the simulations at $\rm t=t_{ff}$. Left panel shows the IMF for simulation A, and the right panel shows the IMF for simulation B. The green dashed line represents the Salpeter function. The blue dot-dashed line represents the Chabrier IMF, as fitted to simulation B. A power-law fit is applied to the data above the turn-over mass, $\sim$1.2$\rm ~M_{\odot}$ left and $\sim$0.7$\rm ~M_{\odot}$ right, and is shown as the solid purple line.}
\label{fig:imf}
\end{figure*}

Simulation A starts forming protostars at around two-thirds of the initial free-fall time, which is later than simulation B, but at about three-quarters of the free-fall time, we see a sudden increase of star formation. Simulation B, on the other hand, has a gradual increase of protostars, starting at about a half t$_{\rm ff}$, only to peak close to the free-fall time of 10$^{5}$ years. In the end, simulation B has created more sink particles, 153 against 118, but their masses are much lower. We plot the particle masses against the number of particles in Fig. \ref{fig:imf}.

The plots show that the X-ray dominated environment clearly has a higher characteristic mass as well as a higher minimum and maximum mass compared to simulation B. Higher sink particle masses are seen consistently throughout simulation A. Another difference is evident when looking at the slope of the IMF, which is defined as

\begin{equation}
\rm \frac{dN}{dM} \propto M^{-\alpha} \Longrightarrow \frac{dlogN}{dlogM} = -\alpha + 1 = \Gamma,
\label{eq:imf}
\end{equation}
with N the number of stars in a range of mass dM, $\alpha$ the power-law index, and $\Gamma$ the slope above the characteristic mass of $\sim$0.3-0.5 M$_{\odot}$.

Simulation A has a much flatter slope than Simulation B, which is $\Gamma_{\rm A}=$ -0.28 after 10$^{5}$ years of evolution, flatter than the Salpeter slope of -1.35 \citep{1955ApJ...121..161S}. There is also a steep decrease in the number of stars above $\sim$3 $\rm M_{\odot}$. This slope is very steep but uncertain and can depend strongly on late accretion and merging \citep{2010MNRAS.tmp..405D}. Simulation B, has a slope of $\Gamma_{\rm B}=$ -1.37 which, in fact, is very similar to the Salpeter slope. Our simulations show that these trends for simulation B are maintained even at higher dynamical times, $\rm t=2t_{ff}$, in agreement with the time independent, competitive accretion driven results of \cite{2008MNRAS.386....3C} for initial energies of $\rm |E_{grav}| \geqslant \rm E_{kin}$. We attribute the difference in $\Gamma$ partly to the higher gas temperatures ($\sim$50 K) caused by X-rays, which in turn increases the Jeans mass and the fragmentation mass scale. Most of the massive sink particles are created in the dense fragments and trace the warm X-ray heated (T$>$50 K) molecular gas. The Jeans mass in case B is about 0.2 M$_{\odot}$ after one free-fall time, whereas the Jeans mass in case A can be much higher, but typically ranges from 0.5 to 2 M$_{\odot}$ in the $\rm \gtrsim 10^{6} ~cm^{-3}$ gas. The Jeans mass is approximately given by \citep{2007A&A...475..263F};

\begin{equation}
\rm M_{J} = \left( \frac {\pi c_s^2} {G} \right)^{\frac{3}{2}} \rho^{-\frac{1}{2}}
\simeq \frac{90} {\mu^{2}} ~T^{\frac{3}{2}} ~n^{-\frac{1}{2}} ~(M_{\odot}).
\label{eq:jeans}
\end{equation}

In response to the high temperature in the XDR gas, only high density regions collapse creating more massive protostars and, in the end, fewer of them. Furthermore, the cloud is shielded from radiation with increasing column density, causing less X-ray penetration and thus less heating, lowering the Jeans mass. As a result, gas becomes more compressible and the density increases, starting a snowball effect. This causes a steep decrease of the Jeans mass, giving rise to a sharp increase in star formation. Another feature of the aforementioned effect is that almost all of the particles that are created in simulation A, are formed in the high density front, the finger head, of the cloud, where the snowball effect is initiated.

\subsection{Sink particle behavior}
We let the code check whether sink particles come too close to one another and if they should merge. This is the case if the velocity difference between two particles is less than the escape velocity and if the merging time is shorter than the simulation time step. Merging is not often seen. We define the merging time as half of a degenerate orbit time, $\rm t_{merge} = \pi R_{12}^{3/2} ~(\rm 8G(m_{1}+m_{2}))^{-1/2}$, where m$_{1}$ and m$_{2}$ are the masses of two independent sink particles and R$_{12}$ is the distance between the two.

Mergers can potentially increase the mass considerably, but most of the protostellar mass gain comes from accretion. To this end, we adopted a Bondi-Hoyle type of accretion \citep{2004ApJ...611..399K}. This type of accretion arises when a homogeneous flow of matter at infinity moves non-radially towards the accretor. Accretion increases with protostellar mass, but drops with decreasing ambient density and higher Mach numbers. We see that the ratio of the most massive to the second most massive sink particle in a local cloud region grows in time. With this, our expectation is confirmed that protostars compete with each other for material and that high mass stars as well as stars that lie in deep potential wells accrete more \citep{2006MNRAS.370..488B, 2008ASPC..390...26B}.

The total mass of the particles after one free-fall time for simulation B is 104 M$_{\odot}$, i.e., about 1/8th of the initial cloud mass. Quite interestingly, simulation A is able to convert more gas mass into stars, 230 M$_{\odot}$. This is a strong argument for efficient X-ray induced star formation in AGN. The star formation efficiencies mentioned here are upper limits, since these calculations do not contain feedback effects from young stars. Feedback effects, like outflows, should decrease the efficiency overall \citep{2010ApJ...709...27W}.

\section{Conclusions and Discussion}
We have performed two 3D simulations of similar molecular clouds, each at 10 pc distance from a supermassive black hole and followed their evolution. In one case (A), we expose the cloud to an active black hole, producing a strong, 160 $\rm erg ~s^{-1} ~cm^{-2}$, X-ray flux. In the other case (B) the black hole was inactive and the molecular cloud had isothermal (10 K) conditions throughout the run. We saw clear differences between the simulations.

For the X-rays included run, we find that the molecular cloud is heated at the irradiated side and an ionizing pressure front is formed. This conic pressure front breaks up due to turbulent motions, forming finger-like structures that can be seen in column density plots, see Fig. \ref{fig:column}. The density increases at the top of the compression front and the shielded gas cools. Star formation is initially delayed due to the higher temperatures, but the cloud continues to contract and after the critical density is reached sink particles start to from. Since the temperature in the shielded parts of the cloud decreases with increasing density, the Jeans mass drops at an amplified rate. Consequently, protostar formation increases sharply around 0.75 $\rm t_{ff}$. Despite this, fewer sink particles are created in total with respect to simulation B, but they are more massive. The latter is a consequence of the high temperature ($\sim$50 K) and Jeans mass ($\gtrsim$1 M$_{\odot}$) in X-ray irradiated gas. In the end, case A ends up with more total stellar mass after one free-fall time. These protostars also accrete more material as the accretion rate scales with mass and density. This becomes increasingly more important for the mass growth due to the deeper potential well created and favored by massive stars. The two effects together cause that competitive accretion dominates the mass growth and strongly affects the shape of the mass function. The resulting IMF has a higher characteristic mass and a near-flat, non-Salpeter slope. We summarize the main points by stating how case A behaves with respect to B:

\begin{itemize}
 \item Protostars are created at a later stage, around 0.65 $\rm t_{ff}$ (0.45 $\rm t_{ff}$ for B).
 \item Fewer protostars are created but they have higher masses, M=0.8-6 M$_{\odot}$ (0.2-3.6 M$_{\odot}$ for B).
 \item Although protostars are created later, there is a burst mode around 3/4th of the collapse time of 10$^{5}$ years.
 \item The total stellar mass formed from the gas is more than twice higher after 10$^{5}$ years and the efficiency is about 28$\%$ (13$\%$ for B).
 \item Competitive accretion is more influential in shaping the IMF.
 \item The IMF has a much flatter slope, $\Gamma=-0.28$ (-1.37 for B).
 \item The characteristic mass of the IMF is higher, $\rm M_{char} \sim$1.2 $\rm M_{\odot}$ ($\sim$0.7 $\rm M_{\odot}$ for B).
\end{itemize}

All of these results tell us two main things. Firstly, star formation is induced by X-rays and massive stars can form in this way with high efficiency. Secondly, the resulting IMF differs from a Salpeter shape.

The pressure force exerted by radiation from the black hole is  found to be modest. The bulk of the radiation pressure will results from the UV, since it dominates the bolometric luminosity L for very massive black holes. Typically, the change in momentum of the gas is proportional to $\rm \tau_{\rm FIR} ~L/c$, for the speed of light c and mean FIR optical depth $\tau_{\rm FIR}$. The latter comes in because the dust, if it absorbs the bulk of the hard radiation, re-emits this in the FIR (100-300 $\rm \mu m$). The ratio of radiation pressure over thermal pressure, for a typical $\tau_{\rm FIR}$ of $\sim$0.1, is about 10\% at 10 pc from the black hole

We did not include any model for stellar feedback in this study. The total time of the simulations is not long enough for stellar evolution effects to play a major role, however, outflows from young stars can produce significant winds that disrupt gas build-up. This may affect the later time ($\rm >1~t_{ff}$) accretion onto protostars.

There is also strong gravitational shear, which helps to drive the turbulence but also compresses some regions in the center of the cloud along the direction perpendicular to the orbit. This increases the star formation efficiency overall \citep{2008Sci...321.1060B}, but does not affect the conclusions when comparing the two cases with each other. 

There are several other parameters that can be studied for their effectiveness, like the X-ray flux, turbulent strength, cloud size and initial gas density, distance to, and mass of, the black hole. In a follow-up paper we plan to present a detailed parameter study to this effect including the impact of UV and cosmic rays.

\begin{acknowledgements}
We are very grateful to the anonymous referee for an insightful and constructive report that greatly helped this work. The FLASH code was in part developed by the DOE-supported Alliance Center for Astrophysical Thermonuclear Flashes (ACS) at the University of Chicago. Part of the simulations have been run on the dedicated special purpose machines `Gemini' at the Kapteyn Astronomical Institute, University of Groningen. We are also grateful for the subsidy and time granted, by NCF, on the `Huygens' national supercomputer of SARA Amsterdam to run our high resolution simulations.
\end{acknowledgements}

\bibliography{AstroPH-HocukSpaans2010a.bib}

\end{document}